\documentstyle[12pt,equation,axodraw]{article}
\begin{document}
\newcommand{\be}{\begin{equation}}
\newcommand{\ben}{\begin{subequations}}
\newcommand{\een}{\end{subequations}}
\newcommand{\beq}{\begin{eqalignno}}
\newcommand{\eeq}{\end{eqalignno}}
\newcommand{\ee}{\end{equation}}
\newcommand{\wt}{\widetilde}
\newcommand{\stw}{\sin^2 \! \theta_W}
\newcommand{\ctw}{\cos^2 \! \theta_W}
\newcommand{\sw}{\sin \! \theta_W}
\newcommand{\cw}{\cos \! \theta_W}
\newcommand{\dmchi}{\mbox{$\Delta m_{\tilde {\chi}}$}}
\newcommand{\mchi}{\mbox{$m_{\tilde {\chi}_1^0}$}}
\newcommand{\mchisq}{m_{\tilde {\chi}_1^0}^2}
\newcommand{\dtm}{\mbox{${\cos(2\beta) M_Z^2}$}}
\newcommand{\lsp}{\mbox{$\tilde {\chi}_1^0$}}
\newcommand{\Ochi}{\mbox{$\Omega_{\tilde \chi} h^2$}}
\newcommand{\tanb}{\mbox{$\tan \! \beta$}}
\newcommand{\cotb}{\mbox{$\cot \! \beta$}}
\newcommand{\cosb}{\mbox{$\cos \! \beta$}}
\newcommand{\sinb}{\mbox{$\sin \! \beta$}}
\newcommand{\ghcc}{\mbox{$g_{h\tilde{\chi}_1^0 \tilde{\chi}_1^0}$}}
\newcommand{\gHcc}{g_{H\tilde{\chi}_1^0 \tilde{\chi}_1^0}}
\newcommand{\gAcc}{g_{A\tilde{\chi}_1^0 \tilde{\chi}_1^0}}
\newcommand{\gZcc}{\mbox{$g_{Z\tilde{\chi}_1^0 \tilde{\chi}_1^0}$}}
\newcommand{\thb}{\theta_{\tilde b}}
\newcommand{\tht}{\theta_{\tilde t}}
\newcommand{\thq}{\theta_{\tilde q}}
\newcommand{\sto}{\tilde{t}_1}
\newcommand{\sbo}{\tilde{b}_1}
\newcommand{\sqo}{\tilde{q}_1}
\newcommand{\stt}{\tilde{t}_2}
\newcommand{\sbt}{\tilde{b}_2}
\newcommand{\sqt}{\tilde{q}_2}
\renewcommand{\thefootnote}{\fnsymbol{footnote}}

\begin{flushright}
APCTP 96--06 \\
KEK--TH 505 \\
TIFR/TH/96--62 \\
TU--515 \\
December 1996\\
\end{flushright}

\vspace*{2.5cm}
\begin{center}
{\Large \bf  Light Higgsino Dark Matter} \\
\vspace{10mm}
Manuel Drees$^1$, Mihoko M. Nojiri$^2$, D.P. Roy$^3$ and Youichi Yamada$^4$ \\
\vspace{5mm}
${}^1${\it APCTP, College of Natural Sciences, Seoul National University, 
Seoul 151--742, Korea} \\
${}^2${\it KEK Theory Group, Oho-ho 1-1, Tsukuba, Ibaraki, 305 Japan} \\
${}^3${\it Tata Institute of Fundamental Research, Homi Bhabha Road, 
Mumbai 400005, India} \\
${}^4${\it Department of Physics, Tohoku University, Sendai, 980-77, 
Japan}

\end{center}
\vspace{10mm}

\begin{abstract}
We re--investigate the question whether a light higgsino--like
neutralino is a viable Dark Matter candidate. To this end we compute
the dominant one--loop corrections to the masses of the higgsino--like
states in the minimal Supersymmetric Standard Model (MSSM), due to
loops involving heavy quarks and their superpartners. We also
calculate analogous corrections to the couplings of higgsino--like
neutralinos to $Z$ and Higgs bosons. In the region of parameter space
leading to high higgsino purity of the lightest neutralino, these
corrections can change the expected relic density by up to a factor of
five in either direction. We conclude that for favorable choices of
soft supersymmetry breaking parameters, a state with more than 99\%
higgsino purity could indeed form all cold Dark Matter in the
Universe. In some cases these corrections can also increase the
expected cross section for LSP scattering off spinless nuclei by up to
two orders of magnitude, or reduce it to zero.

\end{abstract}
\clearpage
\setcounter{page}{1}
\pagestyle{plain}
\section*{1) Introduction}
It has been known for more than ten years that the lightest supersymmetric
particle (LSP) can be a good candidate for the missing ``Dark Matter'' (DM)
in the Universe \cite{1,2}. In models with conserved ``R parity'' the
LSP is absolutely stable. Searches for exotic isotopes \cite{3} then imply
that it must be electrically and color neutral. Within the particle content
of the Minimal Supersymmetric Standard Model (MSSM) this leaves us with two
kinds of candidates, the lightest sneutrino $\tilde{\nu}$ and the lightest
neutralino \lsp. A combination of ``new physics'' searches at the CERN
$e^+e^-$ collider LEP and direct DM search experiments excludes the
sneutrino as viable DM candidate \cite{2}. This leaves us with the lightest
neutralino.

In general the lightest neutralino \lsp\ is a superposition of the
$U(1)_Y$ gaugino $\wt{B}$ (``bino''), the neutral $SU(2)$ gaugino
$\wt{W}_3$, and the two higgsinos $\tilde{h}_1^0$ and $\tilde{h}_2^0$
(with $Y_{\tilde{h}_1} = -Y_{\tilde{h}_2} = -1/2$). A
gaugino--dominated state (photino or bino) has the right relic density
\cite{4,5} to provide the missing Dark Matter if
$m^4_{\tilde{l}_R}/m^2_{\tilde{\chi}_1^0} \simeq (200 \ {\rm GeV})^2$,
which points towards the same range of superparticle masses favored by
naturalness arguments; here $\tilde{l}_R$ stands for $SU(2)$ singlet
sleptons, whose exchange dominates the annihilation of bino--like neutralinos
since they have the largest hypercharge. Originally it was thought \cite{4}
that a sufficiently pure higgsino--state would also give an interesting relic
density, if it is lighter than the $W$ boson so that $\lsp \lsp \rightarrow
W^+ W^-$ is kinematically suppressed. The reason is that the coupling of
a pair of higgsino--like LSPs to a $Z$ boson becomes very small if the
gaugino masses are much larger than the higgsino mass parameter. However,
Mizuta and Yamaguchi \cite{6} later realized that the standard estimate \cite{2}
for the LSP density is not reliable in this case. The reason is that there are
actually three higgsino--like states, two neutralinos and one chargino. The
mass splitting between these states becomes very small as the gaugino masses
become large. In such a situation one has to include ``co--annihilation''
between the LSP and these only slightly heavier states \cite{7}. Note that
the $Z \lsp \tilde{\chi}_2^0$ and $W \lsp \tilde{\chi}_1^\pm$
couplings are large if the LSP is dominantly a higgsino. Co--annihilation
therefore greatly reduces the estimate for the relic density of a
higgsino--like LSP \cite{5,6}.

So far the discussion was based essentially on tree--level results (although
QCD corrections were taken into account in the leading logarithmic approximation
by using ``running'' quark masses and couplings \cite{5}). More recently,
complete one--loop electroweak radiative corrections to the masses
of the neutralinos and charginos have become available \cite{8}.
Using these general results, Giudice and Pomarol very recently pointed out
\cite{9} that loop corrections can quite significantly change the mass
splitting between the three higgsino--like states, the dominant contribution
coming from loops involving heavy quarks ($t,b$) and their superpartners.
This is of some relevance for superparticle searches at LEP, since the
heavier higgsino--like states will be quite difficult to detect
experimentally if their decays only deposit a few GeV of visible energy in
the detector.\footnote{In this case one can still search for events where
the heavier higgsinos are produced in association with a hard, isolated
photon. This signal should be viable \cite{10} even in the limit of
almost perfect mass degeneracy, but the cross section is considerably
smaller than for the simple pair production process.}

Here we point out that these radiative corrections can also change the
estimate of the LSP relic density quite dramatically, since the 
co--annihilation rate depends exponentially on the mass splitting
between the higgsino--like states. Radiative corrections also change
the decomposition of the LSP, which alters its couplings to gauge and Higgs
bosons. These couplings are further modified by explicit vertex corrections.
We present a full calculation of these three--point function corrections
due to Yukawa interactions. Their effect on the relic density is
relatively modest, but for negative sign of the higgsino mass parameter
they can change the cross section for LSP scattering off spinless nuclei
by two orders of magnitude.

The remainder of this article is organzied as follows. In Sec.~2 we
describe the formalism, including one--loop corrections to the masses
and relevant couplings of higgsinos. In Sec.~3 we present numerical
results for the LSP relic density and its detection rate in a
$^{76}$Ge detector. Our estimate of the relic density includes a
careful treatment of $s-$channel poles \cite{7}, as well as
``sub--threshold annihilation'' \cite{7} into $W$ and Higgs boson
pairs. Our estimate of the LSP detection rate includes the full set of
contributions discussed in ref.\cite{11}. In Sec.~4 we summarize our 
results and present some conclusions. Finally, in the Appendix we list
expressions for three--point functions in the kinematical configurations of
interest to us.

\section*{2) Formalism}
In this section we describe the calculation of the one--loop corrections to
the mass of the higgsino--like states, as well as to the couplings of the
LSP to $Z$ and Higgs bosons. We focus here on corrections from Yukawa
couplings, which give the potentially largest contributions to the mass
splittings \cite{9} and couplings of interest to us.

\subsection*{2a) Corrections to the Masses}
The corrections to the mass splittings can be understood as corrections to the
chargino and neutralino mass matrices. Including one--loop corrections to
the higgsino masses, these matrices can be written as
\ben \label{e1} \beq 
{\cal M}_\pm &= \mbox{$ \left( \begin{array}{cc}
M_2 & \sqrt{2} M_W \sin \! \beta \\
\sqrt{2} M_W \cos \! \beta & \mu + \delta_C
\end{array} \right) $} \label{e1a}\\
{\cal M}_0 &= \mbox{$ \left( \begin{array}{cccc}
M_1 & 0 & - M_Z \cos \! \beta \sin \! \theta_W & M_Z \sin \! \beta \sin \!
\theta_W \\
0 & M_2 &   M_Z \cos \! \beta \cos \! \theta_W & -M_Z \sin \! \beta \cos \!
\theta_W \\
- M_Z \cos \! \beta \sin \! \theta_W & M_Z \cos \! \beta \cos \! \theta_W & 
\delta_{33} & -\mu - \delta_{34} \\
M_Z \sin \! \beta \sin \! \theta_W & -M_Z \sin \! \beta \cos \! \theta_W &
-\mu - \delta_{34} & \delta_{44}
\end{array} \right) $} \label{e1b}
\eeq \een
Here $M_1$ and $M_2$ are the supersymmetry breaking masses of the $U(1)_Y$ and
$SU(2)$ gauginos, respectively, $\mu$ is the higgsino mass parameter, and
$\tanb = \langle H_2^0 \rangle / \langle H_1^0 \rangle$ is the ratio of
vacuum expectation values of the two neutral Higgs fields of the MSSM
\cite{15,18}.

Note that we are only interested in states that are (almost) pure higgsinos.
We do therefore not include contributions that are suppressed by both loop
factors and small higgsino--gaugino mixing angles. This is why we only 
included corrections to the higgsino entries of the mass matrices. Further,
since we focus on corrections due to Yukawa couplings, we only need to 
consider quark--squark loops. Their contributions are given by \cite{8}:
\ben \label{e2} \beq
\delta_C &= - \frac{3} {32 \pi^2} \Re e\left\{ h_b h_t
\sin(2\thb) m_t \left[ B_0(Q,t,\sbo) - B_0(Q,t,\sbt) \right] 
\right. \nonumber \\ & \left.
\hspace*{2cm} + h_b h_t \sin(2\tht) m_b \left[
B_0(Q,b,\sto) - B_0(Q,b,\stt) \right]
\right. \\ & \left.
\hspace*{2cm} +  \mu \left[ \left( h_b^2 \sin^2 \thb + h^2_t
\cos^2 \thb \right) B_1(Q,t,\sbo) + \left( h_b^2 \cos^2 \thb + h_t^2 \sin^2
\thb \right) B_1(Q,t,\sbt)
\right. \right. \nonumber \\ & \left. \left. \hspace*{2.35cm}
+ \left( h_t^2 \sin^2 \tht + h_b^2 \cos^2 \tht \right) B_1(Q,b,\sto)
+ \left( h_t^2 \cos^2 \tht + h_b^2 \sin^2 \tht \right) B_1(Q,b,\stt)
\right] \right\}; 
\label{e2a} \nonumber \\
\delta_{34} &= - \frac {3\mu} {32 \pi^2} \Re e\left\{ h_t^2 \left[
B_1(Q,t,\sto) + B_1(Q,t,\stt) \right] 
+ h_b^2 \left[ B_1(Q,b,\sbo) + B_1(Q,b,\sbt) \right] \right\};
\label{e2b} \\
\delta_{33} &= - \frac {3}{16 \pi^2} h_b^2 m_b \sin(2\thb) \Re e\left\{
B_0(Q,b,\sbo) - B_0(Q,b,\sbt) \right\}; \label{e2c} \\
\delta_{44} &= - \frac{3}{16 \pi^2} h_t^2 m_t \sin(2\tht) \Re e\left\{
B_0(Q,t,\sto) - B_0(Q,t,\stt) \right\}. \label{e2d}
\eeq \een
Here $B_0$ and $B_1$ are two--point functions, for which we use the conventions
of refs.\cite{8}. Their first argument is the external momentum scale $Q$,
and the second and third arguments are a quark and squark mass, for which
we wrote the symbol of the corresponding fields in order
to avoid double subscripts.
The squark masses are eigenvalues of the $\tilde{t}$ and $\tilde{b}$
mass matrices \cite{12}, which we write in the basis $(\tilde{q}_L,
\tilde{q}_R)$ following the notation of ref.\cite{13}:
\ben \label{e3} \beq
{\cal M}^2_{\tilde t} &= \mbox{$ \left( \begin{array}{cc}
m_t^2 + m^2_{\tilde{t}_L} + \left( \frac{1}{2} - \frac{2}{3} \stw \right)
\dtm & - m_t \left( A_t + \mu \cotb \right) \\
- m_t \left( A_t + \mu \cotb \right) & m_t^2 + m^2_{\tilde{t}_R} + 
\frac{2}{3} \stw \dtm
\end{array} \right) $} \label{e3a} \\
{\cal M}^2_{\tilde b} &= \mbox{$ \left( \begin{array}{cc}
m_b^2 + m^2_{\tilde{t}_L} - \left( \frac{1}{2} - \frac{1}{3} \stw \right)
\dtm & - m_b \left( A_b + \mu \tanb \right) \\
- m_b \left( A_b + \mu \tanb \right) & m_b^2 + m^2_{\tilde{b}_R} - 
\frac{1}{3} \stw \dtm
\end{array} \right) $} \label{e3b}
\eeq \een
Note that $SU(2)$ invariance implies that the soft breaking terms appearing
in the (1,1) entries of the mass matrices (\ref{e3a}) and (\ref{e3b}) are 
equal. 
The lighter eigenstates are defined as $\sqo = \tilde{q}_L \cos \thq + 
\tilde{q}_R \sin \thq$. Finally, $h_b$ and $h_t$ in eqs.(\ref{e2}) are the
Yukawa couplings of the $b$ and $t$ quarks:
\be \label{e4}
h_b = \frac {g m_b} { \sqrt{2} M_W \cosb}; \hspace*{2cm}
h_t = \frac {g m_t} { \sqrt{2} M_W \sinb}, 
\ee
where $g$ is the $SU(2)$ gauge coupling, and the quark masses are to be
taken at scale $Q$.

As written, the corrections $\delta_C$ and $\delta_{34}$ are divergent. The
fact that the divergence is the same for these two quantities provides a
nontrivial check of our calculation. This divergence has to be absorbed by
renormalizing the higgsino mass parameter $\mu$. We have used the 
$\overline{\rm DR}$ renormalization scheme, with renormalization scale
taken equal to the external momentum scale $Q$. For consistency, the
tree--level parameter $\mu$ in eqs.(\ref{e1}) then has to be interpreted 
as running mass taken at the same scale $Q$. In principle one has to
diagonalize the matrices of eqs.(\ref{e1}) at different $Q=m_{\tilde{\chi}_i^0}$
or $Q=m_{\tilde{\chi}_i^\pm}$ in order to compute the physical (on--shell)
neutralino and chargino masses. However, since the $Q-$dependence of the
corrections is quite weak, for our purposes it is sufficient to compute the
mass matrices at fixed $Q = |\mu|$.

For our later discussion it is convenient to have approximate analytical
expressions for the masses of the higgsino--like states as well as for the
LSP eigenvector. In the for us relevant limit $M_1, M_2 \gg |\mu|$ the
mass of the lighter chargino is approximately
\be \label{e5}
m_{\tilde{\chi}_1^\pm} \simeq | \mu_C| \left[ 1 - \frac {M_W^2 \sin(2\beta)}
{\mu_C M_2} \right] + {\cal O} (M_2^{-2}),
\ee
where $\mu_C = \mu + \delta_C$; see eq.(\ref{e1a}).

In the same limit the two lightest neutralinos are approximately equal to the 
symmetric and anti--symmetric combination of the two higgsino current
eigenstates $\tilde{h}_1^0$ and $\tilde{h}_2^0$. Including terms up to first
order in small quantities, their eigenvectors are given by:
\footnote{In our convention the neutralino
eigenvectors are real, and we keep the signs of the eigenvalues.}
\ben \label{e6} \beq
N_{\tilde{h}_S^0} &\simeq \left( \epsilon_1^{(S)}, \epsilon_2^{(S)}, \frac{1}
{\sqrt{2}} + \epsilon_3, \frac {1} {\sqrt{2}} - \epsilon_3 \right);
\label{e6a} \\
N_{\tilde{h}_A^0} &\simeq \left( \epsilon_1^{(A)}, \epsilon_2^{(A)}, \frac{1}
{\sqrt{2}} - \epsilon_3, - \frac {1} {\sqrt{2}} - \epsilon_3 \right),
\label{e6b}
\eeq \een
with
\ben \label{e7} \beq
\epsilon_1^{(S,A)} &= \frac {M_Z \sw} {M_1} \cdot
 \frac {\cosb \mp \sinb} {\sqrt{2}};
\label{e7a} \\
\epsilon_2^{(S,A)} &= -\frac {M_Z \cw} {M_2} \cdot
\frac {\cosb \mp \sinb} {\sqrt{2}};
\label{e7b} \\
\epsilon_3 &= \frac {M_Z^2 \cos(2\beta)} {4 \sqrt{2} \mu_N} \left( 
\frac{\stw}{M_1} + \frac{\ctw}{M_2} \right) + \frac {\delta_{44}-\delta_{33}}
{4 \sqrt{2} \mu_N} \nonumber \\
&= \frac {\sqrt{2} M_W^2 \cos(2\beta)} {5 \mu_N M_2}
+ \frac {\delta_{44}-\delta_{33}} {4 \sqrt{2} \mu_N},
\label{e7c}
\eeq \een
with $\mu_N = \mu + \delta_{34}$. The upper (lower) sign in eqs.(\ref{e7}a,b)
holds for the symmetric (anti--symetric) higgsino state. The second equality
in eq.(\ref{e7c}) is valid only if one assumes the usual 
``unification condition''
\be \label{e8}
M_1 = \frac{5}{3} \tan^2 \! \theta_W M_2 \simeq 0.5 M_2.
\ee

The masses of the higgsino--like eigenstates are given by
\beq \label{e9}
m_{\tilde{h}^0_{S,A}} &\simeq \mp \mu_N - \frac {M_Z^2}{2} \left( 1 \mp 
\sin(2\beta) \right) \left( \frac{\stw}{M_1} + \frac{\ctw}{M_2} \right)
+ \frac {1}{2} \left( \delta_{33} + \delta_{44} \right) 
\nonumber \\
&\simeq \mp \mu_N - \frac {4 M_W^2} {5 M_2} \left( 1 \mp \sin(2\beta) \right)
+ \frac {1}{2} \left( \delta_{33} + \delta_{44} \right) ,
\eeq
where we have kept the signs of the eigenvalues, and the second equality 
again assumes eq.(\ref{e8}).

If the second term in eq.(\ref{e9}) is larger than the loop corrections
given by the third term, which is generally the case for $M_2 \leq 1$ TeV,
the LSP will be the symmetric (anti--symmetric) higgsino state if $\mu$ is
negative (positive). For small and moderate values of \tanb\ this distinction
is quite important, since the anti--symmetric higgsino--like state has
larger gaugino components, see eqs.(\ref{e7}a,b). Moreover, the mass
splitting between the higgsino--like states also depends on the sign of $\mu$.
Assuming for simplicity eq.(\ref{e8}) to hold, one has \cite{9}
\ben \label{e10} \beq
\left| m_{\tilde{\chi}_2^0} - \mchi \right| &\simeq \left| \frac {8 M_W^2}
{5 M_2} - \delta_{33} - \delta_{44} \right|; \label{e10a} \\
m_{\tilde{\chi}_1^\pm} - \left| \mchi \right| &\simeq 
\frac {M_W^2} {5 M_2} \left[ 4 -
{\rm sign}(\mu) \sin(2\beta) \right] - \frac {1}{2} \left( \delta_{33}
+ \delta_{44} \right) + \delta_C - \delta_{34}.
\label{e10b}
\eeq \een
Note that the one--loop corrections increase (decrease) the mass splittings
if $\delta_{33}+\delta_{44}$ is negative (positive). As already pointed out
in ref.\cite{9}, this correction can be quite significant. In contrast, we
found that the last term in eq.(\ref{e10b}), $\delta_C - \delta_{34}$, is
negligible in all cases.

\subsection*{2b) Corrections to the Couplings}
At tree level the $Z \lsp \lsp$ coupling is proportional to the tree--level
contribution to
$\epsilon_3$; this coupling largely determines the annihilation rate
of higgsino--like LSPs. However, at one--loop level one has to include
the explicit vertex correction diagrams of Figs.~1a,b as well as the
(off--diagonal) wave function renormalization diagram of Fig.~1c. We
compute these corrections in the limit where the lighter neutralinos
are exact higgsino; at the end we include mixing by simply multiplying
the correction with the relevant higgsino component of the
eigenstates. This procedure greatly simplifies the calculation. In this
limit the only non--vanishing $Z-$neutralino coupling is the off--diagonal
$Z \lsp \tilde{\chi}^0_2$ coupling, so we do not need to include any
diagonal wave function renormalization diagrams. Note also that there is
no vertex counter--term, since Yukawa couplings do not renormalize gauge
couplings at one--loop level. Our procedure
will give reliable results as long as the gaugino components
$\epsilon_1, \ \epsilon_2$ in eqs.(\ref{e6}) are small. When these
components become sizable or even dominant, our procedure may no
longer give an accurate estimate of the loop corrections; however, in
this case the loop corrections are in any case much smaller than the
tree--level contributions to the mass splittings and couplings of
interest to us, so that we again only make a small error.

\begin{center}
\begin{picture}(400,180)(0,0)
\DashLine(0,90)(50,90){2}
\Line(50,90)(100,140)
\Line(50,90)(100,40)
\DashLine(80,120)(80,60){2}
\Text(15,80)[]{$Z,\Phi$}
\Text(65,115)[]{$q$}
\Text(65,65)[]{$q$}
\Text(88,90)[]{$\tilde{q}_i$}
\Text(105,150)[]{$\lsp(k_1)$}
\Text(105,30)[]{$\lsp(k_2)$}
\Text(50,10)[]{\large a)}
\DashLine(140,90)(190,90){2}
\DashLine(190,90)(220,120){2}
\DashLine(190,90)(220,60){2}
\Line(220,120)(240,140)
\Line(220,60)(240,40)
\Line(220,120)(220,60)
\Text(155,80)[]{$Z,\Phi$}
\Text(205,115)[]{$\tilde{q}_i$}
\Text(205,60)[]{$\tilde{q}_j$}
\Text(228,90)[]{$q$}
\Text(245,150)[]{$\lsp(k_1)$}
\Text(245,30)[]{$\lsp(k_2)$}
\Text(190,10)[]{\large b)}
\DashLine(280,90)(330,90){2}
\Line(330,90)(380,140)
\Line(330,90)(380,40)
\DashCArc(355,115)(12,45,225){2}
\Text(295,80)[]{$Z$}
\Text(359,109)[]{$q$}
\Text(355,136)[]{$\tilde{q}_i$}
\Text(385,150)[]{$\lsp(k_1)$}
\Text(385,30)[]{$\lsp(k_2)$}
\Text(333,106)[]{$\tilde{\chi}_2^0$}
\Text(330,10)[]{\large c)}
\end{picture}
\end{center}

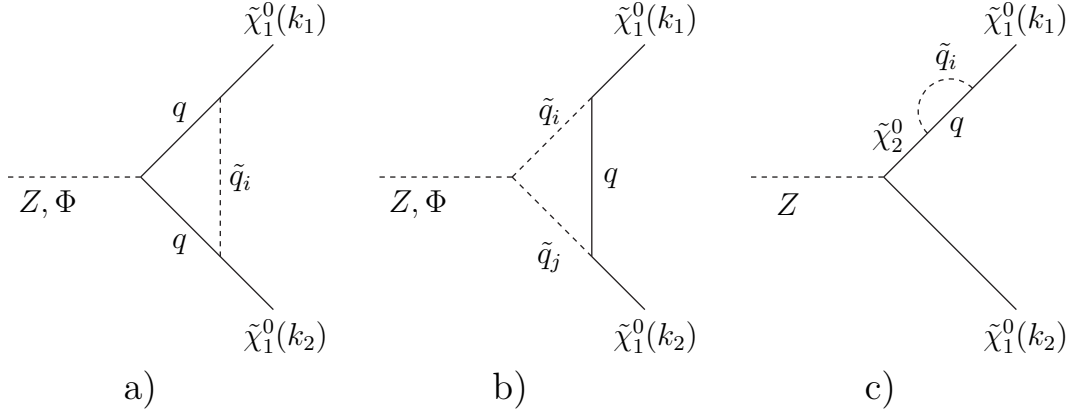
{\bf Fig.~1:} {\small Quark--squark loop corrections to the coupling of a pair
of LSPs to a $Z$ or Higgs boson. The LSP momenta $k_1$ and $k_2$ point
towards the vertex. Note that both senses of the ``Dirac arrow'' (flow of
fermion number) have to be added, since the LSP is a Majorana fermion.
There is also a diagram of type c) with a quark--squark bubble on the
other neutralino line. There are two squark mass eigenstate with
a given flavor.}\\
\vspace*{1cm}

The diagrams of Fig.~1 can be described by the following effective
$Z \lsp \lsp$ vertex:
\be \label{e11}
\delta \Gamma^\mu_{Z \tilde{\chi} \tilde{\chi}} = -i \frac {3 g} {8 \pi^2 \cw}
\left[ \left( N_{13}^2 h_b^2 \delta_a^{(b)}
+ N_{14}^2 h_t^2 \delta_a^{(t)} \right) \gamma^\mu \gamma_5 +
\left( N_{13}^2 h_b^2 \delta_p^{(b)} + N_{14}^2 h_t^2 \delta_p^{(t)} \right)
\left( k_1^\mu + k_2^\mu \right) \gamma_5 \right],
\ee
where $N_{13}$ and $N_{14}$ are the third and fourth component of the LSP
eigenvector, and $k_1$ and $k_2$ are the momenta of the two neutralinos.
We use the tensor decomposition of the three--point function as given in
ref.\cite{14}; this form is convenient for the case of two equal external
masses. The coefficients $\delta_{a,p}$ of eq.(\ref{e11}) can then be
written as:
\ben \label{e12} \beq
\delta_a^{(q)} &= \left[ c_{a,q} + c_{v,q} \cos(2\thq) \right] 
\left[ \left( \mchisq -
k_1 \cdot k_2 \right) C_2^+(\sqo) + \left( \mchisq + k_1 \cdot k_2 \right)
C_2^-(\sqo) 
\right. \nonumber \\ 
& \left. \hspace*{4cm}
+ C_2^0(\sqo) + \mchisq \left( \frac{1}{2} C_0(\sqo) - 2 C_1^+(\sqo) \right)
+ \frac{1}{2} m_q^2 C_0(\sqt) \right]
\nonumber \\ 
&  + \left[ c_{a,q} - c_{v,q} \cos(2\thq) \right] \left[ \left( \mchisq -
k_1 \cdot k_2 \right) C_2^+(\sqt) + \left( \mchisq + k_1 \cdot k_2 \right)
C_2^-(\sqt) 
\right. \nonumber \\ 
& \left. \hspace*{4cm}
+ C_2^0(\sqt) + \mchisq \left( \frac{1}{2} C_0(\sqt) - 2 C_1^+(\sqt) \right)
+ \frac{1}{2} m_q^2 C_0(\sqo) \right]
\nonumber \\ 
& 
- \frac{1}{2} c_{a,q} + m_q \mchi c_{a,q} \sin(2\thq) \left[ C_0(\sqo)
-2 C_1^+(\sqo) - C_0(\sqt) + 2 C_1^+(\sqt) \right]
\nonumber \\ 
& - \cos(2\thq) \left[ \left( -I_{3,q} \cos^2 \! \thq + e_q \stw \right)
C_2^0(\sqo,\sqo) + \left( I_{3,q} \sin^2 \! \thq - e_q \stw \right)
C_2^0(\sqt,\sqt) \right]
\nonumber \\ &
+ I_q \sin^2(2\thq) C_2^0(\sqo,\sqt) - \frac{I_q}{2} \left[ B_1(\sqo)
+ B_1(\sqt) \right]
\nonumber \\ &
+ \frac {m_q I_q} {2 m_{\tilde{\chi}_1^0}} \sin(2\thq) \left[ B_0(\sqt) - B_0(\sqo)
\right] 
\label{e12a} \\
\delta_p^{(q)} &= \mchi \left\{ \left[ c_{a,q} + c_{v,q} \cos(2\thq) \right]
\left[ 2 C_2^-(\sqo) - C_1^+(\sqo) \right]
+ \left[ c_{a,q} - c_{v,q} \cos(2\thq) \right]
\left[ 2 C_2^-(\sqt) - C_1^+(\sqt) \right]
\right. \nonumber \\ 
& \left. \hspace*{1.cm}
+ 2 \cos(2\thq) \left[  \left( - I_{3,q} \cos^2 \! \thq + e_q \stw \right)
C_2^-(\sqo,\sqo) 
+ \left( I_{3,q} \sin^2 \! \thq - e_q \stw \right) C_2^-(\sqt,\sqt) \right]
\right. \nonumber \\ 
& \left. \hspace*{1.cm}
- 2 I_q\sin^2(2\thq) C_2^-(\sqo,\sqt) \right\}
\nonumber \\ 
& + m_q \sin(2\thq) \left\{ - c_{a,q} \left[ C_1^+(\sqo) - C_1^+(\sqt) \right]
- I_{3,q} C_1^-(\sqo,\sqt) \right\}. \label{e12b}
\eeq \een
Here we have used the shorthand notation $C_k(\tilde{q}_i) = C_k(s,\mchisq,
m_q,m_q,m_{\tilde{q}_i}), \ C_k(\tilde{q}_i,\tilde{q}_j) = C_k(s,\mchisq,
m_{\tilde{q}_i}, m_{\tilde{q}_j}, m_q)$ and $B_k(\tilde{q}_i) = 
B_k(\mchisq,m_q,m_{\tilde{q}_i})$. Recall that we use the $B_1$ function of
refs.\cite{8}, which differs from that of ref.\cite{14} by an overall sign.
Finally, the couplings in eqs.(\ref{e12}) are given by
\be \label{e13}
c_{a,q} = \frac{1}{2} I_{3,q}; \hspace*{2cm} 
c_{v,q} = - \frac{1}{2} I_{3,q} + e_q \stw,
\ee
where $I_{3,q} = \pm 1/2$ and $e_q$ are the weak isospin and charge of
quark $q$, respectively. 

We note that one obtains a finite result only after summing over all three
classes of diagrams and both squark eigenstates. Similarly, decoupling of
degenerate heavy squarks ($m_{\tilde{q}_1} \simeq m_{\tilde{q}_2} \rightarrow
\infty$) only holds after summation over all three diagrams and both
squark eigenstates. However, the
very last contribution to $\delta_a^{(q)}$, eq.(\ref{e12a}), is by itself
finite and shows the proper decoupling behaviour. In fact, it closely 
resembles the corrections $\delta_{33}$ and $\delta_{44}$ of eqs.(\ref{e2}c,d).
Indeed, these two terms come from the same two--point function diagrams,
see Fig.~1c. We can therefore include these terms either explicitly in 
$\delta_a^{(q)}$, or via the mass matrix corrections $\delta_{33}, \ 
\delta_{44}$, where they change the quantity $\epsilon_3$ given in 
eq.(\ref{e7c}); recall that this quantity determines the ``tree--level''
$Z\lsp\lsp$ vertex \cite{15}:
%
\beq \label{e14}
\Gamma^{\mu, {\rm tree}}_{Z \tilde{\chi} \tilde{\chi}} &= i
\frac {g} {2 \cw} \gamma^\mu \gamma_5 \left( N_{13}^2 - N_{14}^2 \right)
\nonumber \\
&\simeq \pm i \frac{g} {\cw} \sqrt{2} \epsilon_3 \gamma^{\mu} \gamma_5,
\eeq
where the (upper) lower sign is for the symmetric (anti--symmetric)
higgsino--state, i.e. for negative (positive) $\mu$. Together with
eq.(\ref{e7c}) this sign ensures that the sign of the correction to the
$Z \lsp \lsp$ coupling is independent of the sign of $\mu$. This can
also be seen from the last contribution to $\delta_a^{(q)}$, of course,
keeping in mind that the LSP mass is always positive, independent of the
sign of $\mu$, see eq.(\ref{e9}). We find that this term usually gives the
dominant contribution to $\delta_a^{(q)}$; moreover, the $\delta_p^{(q)}$
are usually quite small. One can therefore get a rough estimate of the
size of the loop contributions from the diagrams of Figs.~1 by simply
diagonalizing the mass matrix (\ref{e1b}), including the corrections
$\delta_{33}$ and $\delta_{44}$, and using the ``tree--level'' vertex
of eq.(\ref{e14}). Of course, one must not include this correction both
in the mass matrix and in $\delta_a^{(q)}$.

We also computed one--loop corrections to the LSP coupling to the neutral
Higgs bosons of the MSSM. These couplings are not so important for the
estimate of the relic density, unless $2\mchi$ happens to be very close to
or slightly lower than the mass of one of these Higgs bosons. However,
the exchange of the neutral scalar Higgs bosons often gives the dominant
contribution \cite{16,11} to elastic LSP--nucleus scattering. One therefore
has to know these couplings quite accurately in order to make a reliable 
estimate of the event rate in various experiments that search for relic
neutralinos. This is true both for direct detection experiments, which
search for the recoil of nuclei struck by ambient LSPs, and for indirect
detection experiments that search for neutrinos produced by LSP annihilation
in the center of the Earth or Sun \cite{17,2}; the detection rate of the
indirect search experiments is proportional to the rate with which ambient
neutralinos are captured by the Earth or Sun, which in turn is proportional
to the LSP scattering cross section off ordinary matter. 

At tree--level a pure higgsino state has {\em no} couplings to Higgs bosons;
these couplings originate from the Higgs--higgsino--gaugino interactions
in the supersymmetric Lagrangian. This also implies that the wave function
renormalization diagram of Fig.~1c does not contribute here.\footnote{There 
are wave function renormalization diagrams where an external higgsino is
converted into a gaugino, which then couples to the Higgs boson and the
second higgsino. This gives contributions of order $\frac{3} {16\pi^2}
g^2 h_t m_t/M_2$, which can be interpreted as ${\cal O}(h_t^2)$
corrections to the tree--level coupling, which is of order $g M_W/M_2$.
However, unlike our corrections $\delta_{33}$ and $\delta_{44}$ this only
corrects an entry in the neutralino mass matrix that is already nonzero
at tree--level, and will therefore change the LSP couplings to Higgs bosons
by at most a few percent. In our case these corrections are further
suppressed since we are interested in a higgsino--like LSP, which (for
$\tanb \ne 1$) implies $M_1, M_2 \gg |\mu|$; diagrams with internal
gaugino lines are then suppressed by the large gaugino masses.} We therefore
only have to evaluate the explicit vertex corrections of Figs.~1a,b. 

Their contribution can be described by the effective vertices
\ben \label{e15} \beq
i\delta \Gamma_{\phi \tilde{\chi} \tilde{\chi}} &= -i \frac {3} {16\pi^2}
\left( h_b^2 N_{13}^2 \delta_\phi^{(b)} + h_t^2 N_{14}^2 \delta_\phi^{(t)}
\right); \label{e15a} \\
i\delta \Gamma_{A \tilde{\chi} \tilde{\chi}} &= - \frac {3} {16\pi^2} \gamma_5
\left( h_b^2 N_{13}^2 \delta_A^{(b)} + h_t^2 N_{14}^2 \delta_A^{(t)}
\right), \label{e15b}
\eeq \een
where $\phi$ stands for the light neutral scalar $h^0$ or the heavy neutral
scalar $H^0$ and $A$ is the pseudoscalar Higgs boson. The coefficients in
eqs.(\ref{e15}) can be written as
\ben \label{e16} \beq
\delta_\phi^{(q)} &= \frac {h_q r_\phi^{(q)}} {\sqrt{2}} \left\{ \sin(2\thq)
\left[ \left( m^2_{\tilde{q}_1} + m_q^2 + \mchisq \right) C_0(\sqo) - 4
\mchisq C_1^+(\sqo)
\right. \right. \nonumber \\ 
& \left. \left. \hspace*{3.1cm}
- \left( m^2_{\tilde{q}_2} + m_q^2 + \mchisq \right) C_0(\sqt) + 4
\mchisq C_1^+(\sqt) \right] 
\right. \nonumber \\ 
& \left. \hspace*{1.7cm}
+ 2 \mchi m_q \left[ C_0(\sqo) + C_0(\sqt) - 2 C_1^+(\sqo) - 2 C_1^+(\sqt)
\right] \right\}
\nonumber \\ 
&
+ c^{(\phi)}_{\tilde{q},11} \left[ m_q \sin(2\thq) C_0(\sqo,\sqo) + 2 \mchi
C_1^+(\sqo,\sqo) \right] 
\nonumber \\ 
& + c^{(\phi)}_{\tilde{q},22} \left[ -m_q \sin(2\thq) C_0(\sqt,\sqt) + 2 \mchi
C_1^+(\sqt,\sqt) \right]
+ 2c^{(\phi)}_{\tilde{q},12} m_q \cos(2\thq) C_0(\sqo,\sqt);
\label{e16a} \\
\delta_A^{(q)} &= \frac {h_q r_A^{(q)}} {\sqrt{2}} \left\{ \sin(2\thq)
\left[ \left( m_q^2 + \mchisq - m^2_{\tilde{q}_1} \right) C_0(\sqo)
- \left( m_q^2 + \mchisq - m^2_{\tilde{q}_2} \right) C_0(\sqt) \right]
\right. \nonumber \\ 
& \left. \hspace*{1.7cm}
+ 2 \mchi m_q \left[ C_0(\sqo) + C_0(\sqt) \right] \right\}
\nonumber \\ &
+ 2 c^{(A)}_{\tilde{q},12} \left[ m_q C_0(\sqo,\sqt) + 2 \mchi \sin(2\thq)
C_1^-(\sqo,\sqt) \right].
\label{e16b}
\eeq \een
Here we have used the same notation for the arguments of the $C$ functions
as in eqs.(\ref{e12}). The coefficients $r_\phi^{(q)}$ and $r_A^{(q)}$
describe the Higgs couplings to quarks; they are given by \cite{18}:
\begin{eqnarray} \label{e17}
r_{H^0}^{(t)} &= - \frac {\sin \! \alpha}{\sinb} ; \hspace*{1.2cm}
r_{h^0}^{(t)} = - \frac {\cos \! \alpha}{\sinb} ; \hspace*{1.2cm}
r_A^{(t)} &=- \cotb ; \nonumber \\
r_{H^0}^{(b)} &= - \frac {\cos \! \alpha}{\cosb} ; \hspace*{1.2cm}
r_{h^0}^{(b)} = \frac {\sin \! \alpha}{\cosb} ; \hspace*{1.2cm}
r_A^{(b)} &=- \tanb,
\end{eqnarray}
where $\alpha$ is the mixing angle of the neutral scalar Higgs bosons \cite{18}.
Finally, the coefficients $c^{(\phi)}_{\tilde{q},ij}$ and 
$c^{(A)}_{\tilde{q},12}$ describe the couplings of one Higgs boson to a pair
of squarks; they are given by \cite{19}
\ben \label{e18} \beq
c^{(H^0)}_{\tilde{q},11} &= - \frac {g M_Z \cos(\alpha+\beta)} {\cw}
\left[ I_{3,q} \cos^2 \! \thq - e_q \stw \cos(2\thq) \right]
+ \frac {g m_q^2} {M_W} r_{H^0}^{(q)} 
\nonumber \\ &
- \frac {h_q \sin(2\thq)} {\sqrt{2}}
\left( r_{H^0}^{(q)} A_q + r_{H^0}^{'(q)} \mu \right); 
\label{e18a} \\
c^{(H^0)}_{\tilde{q},12} &= - \frac {g M_Z \cos(\alpha+\beta)} {\cw}
\sin(2\thq) \left[ e_q \stw - \frac{I_{3,q}}{2} \right]
\nonumber\\
& - \frac {h_q \cos(2\thq)} {\sqrt{2}}
\left( r_{H^0}^{(q)} A_q + r_{H^0}^{'(q)} \mu \right); 
\label{e18b} \\
c^{(h^0)}_{\tilde{q},11} &=  \frac {g M_Z \sin(\alpha+\beta)} {\cw}
\left[ I_{3,q} \cos^2 \! \thq - e_q \stw \cos(2\thq) \right]
+ \frac {g m_q^2} {M_W} r_{h^0}^{(q)} 
\nonumber \\ &
- \frac {h_q \sin(2\thq)} {\sqrt{2}}
\left( r_{h^0}^{(q)} A_q + r_{h^0}^{'(q)} \mu \right); 
\label{e18c} \\
c^{(h^0)}_{\tilde{q},12} &= \frac {g M_Z \sin(\alpha+\beta)} {\cw}
\sin(2\thq) \left[ e_q \stw  - \frac{I_{3,q}}{2} \right]
- \frac {h_q \cos(2\thq)} {\sqrt{2}}
\left( r_{h^0}^{(q)} A_q + r_{h^0}^{'(q)} \mu \right); 
\label{e18d} \\
c^{(\phi)}_{\tilde{q},22} &= c^{(\phi)}_{\tilde{q},11} \left( \cos \! \thq
\rightarrow \sin \! \thq, \ \sin \! \thq \rightarrow - \cos \! \thq \right);
\label{e18e} \\
c^{(A)}_{\tilde{q},12} &= \frac {h_q} {\sqrt{2}} \left(  r_A^{(q)} A_q
+ \mu \right)\times \left(\begin{array}{c} 
\sin\beta {\rm \ for \ } u\\
\cos\beta {\rm \ for \  }d\\
\end{array}\right).
\label{e18f}
\eeq \een
Where 
\begin{eqnarray} \label{e19a}
r_{H^0}^{'(t)} &= - \frac {\cos \! \alpha}{\sinb} ; \hspace*{1.2cm}
r_{h^0}^{'(t)} &= \ \frac {\sin \! \alpha}{\sinb} ; \nonumber \\
r_{H^0}^{'(b)} &= - \frac {\sin \! \alpha}{\cosb} ; \hspace*{1.2cm}
r_{h^0}^{'(b)} &= - \frac {\cos \! \alpha}{\cosb}.
\end{eqnarray}

The $A_q$ also appear in the squark mass matrices of eqs.(\ref{e3}), and
$I_{3,q}$ and $e_q$ again refer to the weak isospin and charge of quark $q$.
Finally, the Yukawa couplings $h_q$ have been defined in eq.(\ref{e4}).

Note that the pseudoscalar Higgs boson has no couplings to two equal squark
eigenstates \cite{18}.

We note that in this case the diagrams of the type shown in Fig.~1a are finite
by themselves once one has summed over both squark eigenstates, and the
diagrams of Fig.~1b are separately finite for each combination of squarks in
the loop. Notice also that this last class of diagrams is proportional to
the Higss--squark--squark couplings, which receive contributions from the
$A_q$ parameters; these couplings can become very large \cite{19}.

As noted earlier, we use the $C$ functions of ref.\cite{14}. However, there
is a technical complication. When estimating the LSP relic density we need
to evaluate these functions at $s = 4 \mchisq$, whereas LSP--nucleus
scattering cross sections probe these functions at $s \simeq 0$. The
expressions for the higher $C$ functions given in Appendix C of ref.\cite{14}
contain apparent divergencies in {\em both} these limits. We stress that the
loop functions themselves remain well behaved as $s \rightarrow 4 \mchisq$
or $s \rightarrow 0$; the apparent divergencies in the expressions of
ref.\cite{14} therefore all cancel. In fact, even the standard expression for
the scalar three--point function $C_0$ contains apparent divergencies in the
kinematical configurations of interest to us. In the case of $C_0$ the
necessary cancellation between different terms can still be accomplished
numerically by slightly increasing or reducing $s$. However, for the higher
$C$ functions these cancellations become quite delicate. We therefore
re--evaluated the relevant Feynman integrals for the two cases of interest
to us. In both limiting situations the $C$ functions can be expressed as
combinations of two--point ($B$) functions; all coefficients are now
finite. The relevant expressions are collected in the Appendix.

\section*{3) Results}
We are now in a position to present some numerical results. We focus on light
higgsino--like states, $\mchi < M_W$, since heavier LSPs have very large
annihilation cross sections into $W$ and $Z$ pairs \cite{20,5}. A heavy
higgsino therefore only makes a good cold Dark Matter (CDM) candidate if
its mass exceeds 0.5 TeV. This is already uncomfortably heavy for ``weak
scale'' supersymmetry; for example, assuming gaugino mass unification,
the gluino mass has to be larger than 3 TeV in such scenarios.
In fact, the annihilation cross section into $W^+W^-$ final states is so large
that it can be relevant even if the LSP mass is a little below $M_W$.
Such ``sub--threshold annihilation'' can occur since at freeze--out the LSPs
still have significant thermal energy. We include this effect for $W^+W^-$
and also $h^0h^0$ pairs in our estimate of the LSP relic density, using the
formalism developed in ref.\cite{7}. We also use a careful treatment of
$s-$channel poles ($Z$ and Higgs exchange diagrams); as pointed out in
ref.\cite{7}, the standard expansion in the LSP velocity \cite{1}
breaks down in the vicinity of such poles. We use the numerical method
developed in ref.\cite{21}.

In order to illustrate the effects of the loop corrections on the LSP
couplings to Higgs bosons, we also present results for the LSP counting
rate in an isotopically pure ${}^{76}$Ge detector, assuming a fixed local
LSP mass density of 0.3 GeV/cm$^3$ \cite{22} and a velocity dispersion
of 320 km/sec. Nuclear effects are described by a Gaussian form factor, with
a nuclear radius of 4.1 fm \cite{23}. Of course, we could just as well have
used any other spinless isotope. The scattering rate due to spin--dependent
interactions is affected by the loop corrections to the $Z \lsp \lsp$ coupling, 
but this correction is usually somewhat smaller than that to the LSP--Higgs
couplings. Note that the total scattering cross--section off heavy nuclei is
usually dominated by the spin--independent contribution even if the nucleus
in question does have non--vanishing spin \cite{24}.

Since we are interested in scenarios with rather light LSP, we have to be
careful not to violate any experimental bounds. The most relevant constraints
on the parameters appearing in the tree--level neutralino mass matrix
comes from chargino searches at LEP \cite{25}. Unfortunately these bounds are 
not entirely straightforward to interpret in our case, since the standard
set of experimental cuts used to suppress SM backgrounds becomes quite
inefficient in scenarios with small $\dmchi \equiv m_{\tilde{\chi}^\pm_1}
- \mchi$. Note also that the cross section for the production of
higgsino--like charginos is smaller than for gaugino--like states. We 
interpret the LEP bounds as requiring
\be \label{e19}
m_{\tilde{\chi}^\pm_1} \geq  \mbox{$ \left\{ \begin{array}{cc}
75 \ {\rm GeV}, \ \ \ \ \ & \dmchi \geq 10 \ {\rm GeV} \\
45 \ {\rm GeV}, \ \ \ \ \ & \dmchi < 10 \ {\rm GeV} 
\end{array} \right. $}
\ee
The second bound comes from the measurement of the total width of the
$Z$ boson \cite{26}, and thus holds for any value of \dmchi. We are
aware that this parametrization of the LEP search constraints is only
a crude approximation, but it should be sufficient to illustrate the
effects of the loop corrections.

We have seen in the previous section that these corrections depend on the
details of the stop and sbottom mass matrices. In particular, the corrections
$\delta_{33}$ and $\delta_{44}$ to the neutralino mass matrix are proportional
to $\sin(2\thq)$, see eqs.(\ref{e2}c,d); these corrections also vanish in
the limit of equal masses for squarks of a given flavor. The combination
of these two properties means that the corrections depend sensitively
on the size of the off--diagonal entries of the squark mass matrices (\ref{e3}).
Moreover, the potentially largest correction to the LSP--Higgs coupling,
coming from the diagram of Fig.~1b, directly depends on the $A-$parameters
appearing in the squark mass matrices. Third generation squarks also
contribute to other loop processes. This imposes some constraints even
on combinations of parameters where all squark mass eigenstates lie well
above the direct experimental search limits \cite{27,26}.

In ref.\cite{9} the $\tilde{t}-\tilde{b}$ loop contribution
to the electroweak $\rho$ parameter was emphasized. However, given that a
``new physics'' contribution $\delta \rho \simeq 3 \cdot 10^{-3}$ is not
excluded by present data \cite{26}, we find that other loop corrections lead
to stronger constraints. In particular, loop corrections to the mass of the
light neutral Higgs scalar $h^0$ turn negative when $A_t$ becomes too large
\cite{28}. One important constraint therefore comes from searches for the
MSSM Higgs bosons at LEP \cite{26}.

The constraints we have discussed so far do not depend on the masses of the
other squarks, and are therefore quite model--independent. If we make the
additional simplifying assumption that all explicitly supersymmetry breaking
diagonal squark masses ($m_{\tilde{t}_L}, \ m_{\tilde{t}_R}$ and 
$m_{\tilde{b}_R}$ in eqs.(\ref{e3}), and analogous quantities for the
first and second generation squarks) are equal at the weak scale, we find
that the strongest constraint on the parameters of the stop mass matrix
comes from the recent CLEO measurement \cite{29}
of the branching ratio for inclusive $b \rightarrow s \gamma$ decays:
\be \label{e20}
1 \cdot 10^{-4} \leq B(b \rightarrow s \gamma) \leq 4 \cdot 10^{-4}.
\ee
Since we are studying scenarios with rather light charginos, 
chargino--stop contributions to this partial width can be quite large
\cite{30}; they can be of either sign, depending on the signs of $\mu$
and $A_t$. However, the resulting constraint is more model--dependent: If
one allows some non--universality of soft--breaking squark masses, one
also gets contributions from gluino--squark and neutralino--squark loops
\cite{30}, the size of which depends strongly on the details of the entire
three generation squark mass matrices. For defineteness we will stick to a
scenario with exactly universal soft breaking squark masses, and with
$A_t = A_b \equiv A$, with the understanding that the constraints that 
result from imposing the bounds (\ref{e20}) can be relaxed in slightly more
general models without significantly changing the loop corrections to the
masses and couplings of higgsino--like states.

In Fig.~2 we show the dependence of various quantities relevant to our
subsequent analysis, normalized such that they can be plotted to a common
scale. We fixed $M_2 = 350$ GeV and $\mu = -70$ GeV, which means that the
LSP is a more than 99\% pure higgsino; we define the higgsino fraction as
$1 - {\rm gaugino \ fraction} = 1 - \left( \epsilon_1^2 + \epsilon_2^2
\right)$, see eqs.(\ref{e6}). We chose $\tanb=1.5$ so that the top Yukawa
coupling is close to its upper bound, if one requires it to remain
perturbatively small all the way to the GUT scale; on the other hand,
$b-\tilde{b}$ loops are essentially negligible for such a small value of
\tanb. We took a very large mass (1.5 TeV) for the pseudoscalar Higgs
boson, since this maximizes $m_{h^0}$, and hence minimizes the impact of
the LEP Higgs search bounds. This also means that charged Higgs boson loop
contributions to the $b \rightarrow s \gamma$ partial width are negligible.
Our choice of 430 GeV for the common soft breaking squark mass is again
motivated by our desire to maximize the size of the loop effects, given
the experimental constraints discussed above. Increasing $m_{\tilde q}$
for fixed $A/m_{\tilde q}$ would reduce the ratio of physical stop masses,
which leads to reduced $t-\tilde{t}$ loop corrections. On the other hand,
we cannot increase $A/m_{\tilde q}$ beyond the limits shown in Fig.~2
without violating some experimental bound. Finally, here and in the
subsequent figures we assume gaugino mass unification, eq.(\ref{e8}).

The curves in Fig.~2 terminate at values of $A$ where $m_{h^0}$ falls
below the LEP bound of about 62 GeV; note that $h^0$ is essentially
indistinguishable from the single Higgs boson of the SM if $m_A^2 \gg
M_Z^2$. The two dotted curves show a ``high'' and ``low'' theoretical
estimate for $B(b \rightarrow s \gamma)$, scaled up by a factor
$10^4$. Our estimates are based on a leading order QCD analysis
\cite{30}, which has substantial scale uncertainties \cite{31}; the
band in Fig.~2 corresponds to varying the renormalization scale
between 2.5 and 10 GeV, and also includes uncertainties from CKM
matrix elements etc.\footnote{Very recently an almost--complete
next--to--leading order calculation of $B(b \rightarrow s \gamma)$ in
the framework of the MSSM has appeared \cite{32}; their result falls
within our band.} Notice that the ``low estimate'' can be zero. This
happens if the contribution from sparticle loops is larger than that
from the standard $t-W$ loops and has opposite sign, reversing the
sign of the complete matrix element at scale $M_W$ or $m_t$. 
Renormalization group effects give another contribution from
tree--level $W$ exchange due to operator mixing; this contribution is
not sensitive to any ``new physics''. In the SM this term has the same
sign as the loop matrix element at scale $M_W$, leading to a large QCD
enhancement factor, but in the MSSM these two contributions can
cancel. For fixed renormalization scale this cancellation only happens
at specific points of SUSY parameter space, but perfect cancellation
becomes possible for an entire range of parameters if the
renormalization scale is allowed to vary.

We note that for the given sign of $\mu$, the branching ratio for
inclusive $b \rightarrow s \gamma$ decays tends to be below (above) the
SM prediction if $A$ is negative (positive). In order to be conservative,
we only exclude combinations of parameters where the ``high'' theoretical
estimate is below the lower bound, or the ``low'' estimate is above the
upper bound, given in (\ref{e20}). For the parameters of Fig.~2 this
translates into the constraint $-2.7 \leq A/m_{\tilde q} \leq 2.65$, which is
only slightly stronger than that resulting from Higgs searches at LEP. Within
this region, $\delta \rho_{\tilde{t} \tilde{b}} \leq 2.2 \cdot 10^{-3}$.

The solid curve in Fig.~2 shows \dmchi\ (divided by 5 GeV to fit the scale).
The tree--level prediction for this quantity for the given choice of
$M_2, \ \mu$ and \tanb, 14.5 GeV, is very close to the loop--corrected
value for $A=0$. We see that the corrections can either increase or
decrease the chargino--LSP mass splitting by about 4 GeV before one gets
into conflict with the constraint (\ref{e20}). In this case the loop
corrections therefore only amount to at most 30\%; however, we will see
below that this suffices to change the prediction for the LSP relic density
quite dramatically.

Finally, the long and short dashed curves in Fig.~2 show the (re--scaled)
couplings of the LSP to $Z$ and $h^0$ bosons, respectively; \gZcc\ is defined
as the axial vector coupling at $s=4 \mchisq$, while \ghcc\ is defined at
$s=0$. As in case of
\dmchi, the tree--level predictions for these quantities are very close
to the loop--corrected values at $A=0$. We see that the relative variation
in LSP$-Z$ coupling is larger than that in \dmchi, when $A$ is varied over
its allowed range. Note also the positive correlation between these two
quantities, which reinforces the correlation between small
\dmchi\ and small \gZcc\ that holds for higgsino--like LSPs at tree--level,
see eqs.(\ref{e7c}) and (\ref{e10b}). A similar correlation also holds for
the loop--corrected coupling of the LSP to the light scalar Higgs boson,
as shown by the short dashed curve. However, in this case the tree--level
prediction is very small, $\ghcc \simeq 7.6 \cdot 10^{-3}$. This can
be understood from eqs.(\ref{e7}a,b) and the general expression for this
coupling given in ref.\cite{18}:
\beq \label{e21}
g_{h^0 \tilde{\chi}_1^0 \tilde{\chi}_1^0, {\rm tree}} &\simeq
\frac {1} {\sqrt{2}} \left[ \left( g \epsilon_2^{(S,A)} - g'
\epsilon_1^{(S,A)} \right) ( \sin \! \alpha \pm \cos \! \alpha ) \right]
\nonumber \\
&\simeq \frac {4} {5} ( \sinb \mp \cosb )^2 \frac {g M_W}{M_2},
\eeq
where $g'=g \tan \! \theta_W$ is the $U(1)_Y$ gauge coupling, and the
upper (lower) signs again hold for the symmetric (anti--symmetric)
higgsino; in the second step we have used the ``unification condition''
(\ref{e8}) as well as the relation $\alpha \simeq \beta - \frac{\pi}{2}$,
which holds for $m_A^2 \gg M_Z^2$. For the quite small value of \tanb\
used in Fig.~2 this gives a strong cancellation in the coupling of the
symmetric higgsino--like state, which is the LSP for $\mu < 0$. As a result,
the one--loop correction can easily {\em dominate} over the tree--level
contribution (\ref{e21}). This leads to the behaviour shown in Fig.~2,
where the coupling changes sign at $A \simeq - 0.8 m_{\tilde q}$. 

In Figs.~3a--c we show the chargino--LSP mass splitting, the LSP relic
density, and the LSP detection rate in a ${}^{76}$Ge detector as a
function of the gaugino fraction of the LSP eigenstate. We have again
chosen $m_{\tilde q} = 430$ GeV, $\tanb=1.5$ and a large value of
$m_A$.  Note that, unlike in Fig.~2, the physical LSP mass has been
kept fixed in Figs.~3; the value of 70 GeV chosen here is close to
that which maximizes the estimate of the relic density. Since \mchi\
is kept fixed, both $M_2$ and the tree--level parameter $\mu(|\mu|)$
vary along the curves; e.g., $M_2$ lies between about 150 GeV and 1
TeV, with larger values of $M_2$ corresponding to smaller gaugino
fractions, see eqs.(\ref{e7}a,b).  In order to maximize the loop
effects we have also varied the $A-$parameter slightly. In the region
of relatively small $M_2$, i.e. large gaugino fraction, the light
chargino is somewhat heavier, as shown in Fig.~3a; this reduces the
absolute size of the $\tilde{t} - \tilde{\chi}^\pm$ loop
contributions to the $b \rightarrow s \gamma$ decay amplitude for
fixed $A$, which in turn allows us to go to slightly larger values of
$|A|$ without violating the bounds (\ref{e20}).

We show three curves in each of Figs.~3. The dotted curves labelled
``no loops'' have been obtained by switching off the loop corrections
discussed in Sec.~2. However, we keep quark--squark loop contributions
to the mass matrix of the scalar Higgs bosons \cite{33,28}, as well as
$\tilde{q}$ loop contributions to the $h^0 g g$ coupling and $q-\tilde{q}$
loop contributions (box diagrams) to the LSP--gluon coupling \cite{34,11}.
These corrections depend only weakly on the sign of $A$, however. On the
other hand, the signs of the corrections discussed in Sec.~2 are essentially
fixed by the sign of $A$, as shown in Fig.~2. In Figs.~3 we therefore show
results both for positive (solid) and negative (dashed) $A$, keeping $|A|$
fixed. Since we chose parameters close to those that maximize these loop
corrections, the band between the solid and dashed curves in Figs.~3a,b
roughly indicates the range that can be covered by changing the parameters
of the squark mass matrix, for fixed values of the parameters appearing
in the tree--level chargino and neutralino mass matrices.

The results of Fig.~3a show that loop corrections can change the chargino--LSP
mass splitting by about three to four GeV in either direction, as already
indicated in Fig.~2. Note that the absolute size of this correction is almost
independent of the gaugino fraction. This can be understood from 
eq.(\ref{e10b}), which shows that the tree--level and loop--induced 
contributions to \dmchi\ are independent of each other as long as the LSP
is a higgsino--like state.

The results of Fig.~3b show that in the region of high higgsino purity the
relatively modest loop corrections to \dmchi\ can change the estimate of the
LSP relic density by more than a factor of five. The reason is that here the
relic density is essentially determined by co--annihilation processes
\cite{6}, which depend {\em exponentially} on the mass splitting \cite{7}.
Our calculation includes $\lsp \tilde{\chi}_1^\pm$ co--annihilation
into $f \bar{f}'$ and $W \gamma$ final states through $W$ exchange,
and $\lsp \tilde{\chi}_2^0$ co--annihilation into $f \bar{f}$ final
states through $Z$ exchange, where $f$ stands for any SM fermion other
than the top quark. As is usually done, we show results for the LSP mass
density in units of the critical or closure density, $\Omega_{\tilde{\chi}} 
\equiv \rho_{\tilde{\chi}_1^0}/\rho_c$, multiplied with the square of $h$, the
Hubble constant in units of 100 km/(sec$\cdot$Mpc); a conservative
range for $h$ is $0.4 \leq h \leq 0.9$, with recent measurements
clustering around $0.5-0.6$. One needs $\Ochi \geq 0.02 - 0.03$ if the LSP
is to form the bulk of the galactic Dark Matter haloes, and $\Ochi \geq
0.15$ if the LSP is to form {\em all} CDM in models \cite{36} with mixed
hot and cold Dark Matter. The results of Fig.~3b show that, if $A$ is large
and positive, a 99.9\% pure higgsino state can form galactic haloes, and
a 99.5\% pure higgsino might form all CDM. On the other hand, if $A$ is
large and negative, one will need at least 1\% gaugino fraction, corresponding
to $\epsilon_{1,2} \sim 0.1$, even for the LSP to be able to form galactic
haloes.

Note that the curves in Fig.~3b cross over in the region where the gaugino
fraction exceeds several percent. Here the mass splitting between the
higgsino--like states becomes so large that the relic density is again
determined by the usual $\lsp \lsp$ annihilation processes, which in our case
mostly proceed through virtual $Z$ exchange. We saw in Fig.~2 that the loop
corrections increase (reduce) the $Z \lsp \lsp$ coupling if $A$ is positive
(negative). As a result, the curve for $A>0$ reaches its maximum already
at a rather small gaugino fraction; in this case the LSP can form all CDM
if $180 \ {\rm GeV} \leq M_2 \leq 340$ GeV. In contrast, the curve for
$A<0$ reaches its maximum at larger gaugino fraction; here the LSP can
from all CDM only if $M_2$ falls in the narrow window between 160 and 195
GeV.

In Fig.~3c we show estimates for the LSP detection rate in ${}^{76}$Ge,
ignoring possible energy thresholds and assuming a fixed local LSP mass
density. In this case the loop corrections discussed in Sec.~2 can
increase the tree--level result by more than two orders of magnitude!
This is largely due to the small tree--level value of \ghcc\ for the given
case of a symmetric higgsino--like state, see eq.(\ref{e21}). The
turn--over in the region of sizable gaugino fraction is caused by mixing
with the bino--like neutralino; note that near the end of the curves shown
in Figs.~3, $M_1$ and $|\mu|$ are already quite close to each other, so the
expression (\ref{e7a}) for $\epsilon_1$ is no longer reliable.

We emphasize that in this case the band between the solid and dashed
curves is {\em not} a good estimate of the variation of the expected
counting rate when the parameters of the stop mass matrix are varied,
since for large $|A|$ the loop corrections dominate over the
tree--level contribution to \ghcc, as shown in Fig.~2. The total
scattering rate can be made to vanish {\em exactly} for moderately
negative values of $A$; this is related to the change of sign of
\ghcc\ observed in Fig.~2. Note that the scattering rate can vanish
even for quite moderate values of all sparticle masses. This
illustrates that it is impossible to give a strict lower bound on the
expected LSP detection rate even within the MSSM. Of course, there is
no a priori reason for such a cancellation between tree--level and
one--loop contributions to occur; indeed, over most of the parameter
space the loop corrections {\em increase} the expected event rate.
However, even the most optimistic estimate in Fig.~3c is still several
orders of magnitude below the sensitivity of present experiments \cite{37}.

Recall that we do not rescale \cite{38} the event rate in regions of
parameter space leading to a very small LSP relic density; had we done
so, most of the curve for $A<0$ would have been below the one for
positive $A$, as can be seen from Fig.~3b. Finally, we saw in Fig.~2
that for fixed $|A|$, the absolute size of \ghcc\ is somewhat smaller for
$A<0$ than for $A>0$. We nevertheless find a slightly larger scattering
rate for $A<0$, partly because this also gives a slightly smaller value
for $m_{h^0}$; the $h^0$ exchange contribution to the LSP--nucleon
scattering matrix element scales like $\ghcc/m^2_{h^0}$. Destructive
interference with various squark loop diagrams \cite{11} also plays a 
role here.

In Figs.~4a--c we show results similar to those of Figs.~3, but for
positive sign of the higgsino mass parameter $\mu$. The choices for the
other parameters are very similar to those in Figs.~3, except that $A$
is now fixed along each curve. For this rather small value of \tanb, flipping
the sign of $\mu$ has quite dramatic effects, as already anticipated in our
discussion in Sec.~2. In particular the gaugino fraction of the LSP for
fixed values of $M_2$ and $|\mu|$ has become much larger. Conversely, one
has to go to much higher $M_2$ in order to achieve a given level of higgsino
purity; in Figs.~4, $M_2$ varies between about 0.3 and 1.3 TeV. This also
implies that for given higgsino purity the chargino--LSP mass splitting is
smaller for $\mu>0$ than for $\mu<0$, see eq.(\ref{e10b}). On the other
hand, Fig.~4a shows that the size of the loop contributions to this mass
splitting is essentially independent of the sign of $\mu$, as long as the
gaugino fraction is small. Note that for positive $\mu$, \dmchi\ can be
below 10 GeV for a gaugino fraction as large as 12.5\% ($\epsilon_{1,2}
\sim 0.3$); this will have ramifications for chargino searches at LEP
\cite{9}.

The smaller \dmchi\ for fixed gaugino fraction also implies a greatly
reduced relic density, due to enhanced co--annihilation rates. This is
illustrated in Fig.~4b. The increased importance of co--annihilation
also helps to explain why the curves in this figure do not cross, in
contrast to those in Fig.~3b. Another reason is that in the region of
sizable gaugino fraction, $\epsilon_3$ is smaller for positive $\mu$;
this is due to non--leading, ${\cal O}(M_1^{-2})$ terms not included
in eq.(\ref{e7c}), which become quite important when the gaugino
fraction exceeds several percent. Indeed, for the largest gaugino
fractions shown in Figs.~4, our treatment of the loop corrections to
the couplings of the LSP may no longer be entirely reliable, as
discussed in Sec.~2; however, as anticipated in the same discussion,
the relative importance of the loop effects decreases with increasing
gaugino fraction.

This is also true for the estimated LSP detection rate shown in
Fig.~4c. For the smaller gaugino fractions shown in this figure, we
find that the loop corrections can change the estimate by a factor of
about two in either direction, whereas for large gaugino fraction
the loop effects amount to at most 30\%. We note again that, had we
re--scaled the event rate for scenarios with small LSP relic density,
the loop effects would have been even more important in the region of
high higgsino purity. Finally, note that in the region where the
gaugino fraction exceeds 5\% the counting rate in Fig.~4c exceeds that
in Fig.~3c by almost an order of magnitude. This is due to the much
larger tree--level value of \ghcc\ for the anti--symmetric higgsino--like
state, see eq.(\ref{e21}). Since now the tree--level value exceeds the
loop corrections, we find smaller (larger) counting rates for negative
(positive) values of $A$.

As a final illustration of the effects of the loop corrections to the
masses and couplings of higgsino--like states, we show in Figs.~5a--c
the ``geography'' of the well--known $(M_2, \mu)$ plane in the region
$M_2 \gg |\mu|, \ \mu < 0$. In Fig.~5a these loops have been switched off,
while in Figs.~5b,c they have been included with $A = -2.5 m_{\tilde q}$
and $A = +2.5 m_{\tilde q}$, respectively. In each case the region to the right
of the solid line is excluded by the LEP chargino search limit (\ref{e19}).
The long and short dashed curves are contours of constant LSP relic
density $\Ochi = 0.025$ and 0.1, respectively. The remaining lines are
contours of constant LSP detection rate in a ${}^{76}$Ge detector, measured
in events/(kg$\cdot$day); as before, we have assumed a fixed local LSP
density when calculatimg the counting rate.

We see that, depending on the sign of $A$, loop corrections to the
chargino and neutralino mass matrices can signficantly reduce (Fig.~5b)
or increase (5c) the size of the region that is excluded by chargino 
searches at LEP; this is a direct result of the change in \dmchi\
depicted in Fig.~3a. Similarly, the loop corrections can increase or
decrease the region where the LSP is a good CDM candidate; recall that
$\Ochi \geq 0.025$ is required if LSPs are to form the bulk of galactic
Dark Matter haloes. Finally, we again observe a dramatic change of the
expected LSP detection rate due to radiative corrections. Note that this
rate changes only very slowly in Figs.~5b,c, whereas at tree--level one
expects a significant dependence on $M_2$, see Fig.~5a.

We note in passing that results similar to those displayed in Figs.~3
and 5 can also be obtained in the framework of a recently proposed
\cite{39} model with non--unified gaugino masses and a higgsino--like
LSP. This model attempts a supersymmetric interpretion of an event
with an $e^+e^-$ pair, two hard photons, and missing transverse
momentum $p_T$ reported by the CDF collaboration
\cite{40}. The prospects for detecting
relic neutralinos in this model have recently been studied in
refs.\cite{41}, using tree--level results for the neutralino mass
matrix and LSP couplings.  We expect that loop corrections of the type
discussed here can modify these estimates significantly. However,
searches for additional events with two hard photons and missing $p_T$
failed to find additional candidates \cite{42}, casting doubt on any
supersymmetric interpretation of the single anomalous event. We
therefore do not study this model in any further detail.

\section*{4.) Summary and Conclusions}
In this paper we have presented a calculation of loop corrections involving
Yukawa couplings to the masses and couplings of the higgsino--like states 
of the MSSM. We have found these corrections to be very sensitive to the size
and sign of the soft supersymmetry breaking $A$ parameters. If $|A|$ is large,
the one--loop prediction for the difference of the chargino and LSP masses 
can differ by up to $\sim \pm 4$ GeV from the tree--level estimate; the
loop correction to the $\tilde{\chi}_2^0 - \lsp$ mass difference is about
twice as big. Combinations of parameters leading to even larger corrections
lead to conflicts with the measured value of the branching ratio for
inclusive $b \rightarrow s \gamma$ decays, and/or with the negative outcome
of searches for Higgs bosons at LEP. We also found that for negative sign
of the higgsino mass parameter $\mu$, one--loop corrections to the LSP coupling
to $Z$ and Higgs bosons can be comparable to or even larger than the
tree--level contributions already for quite moderate gaugino masses,
$M_2 \geq 200$ GeV. 

We have illustrated the importance of these loop corrections by computing their
effect on the estimated LSP relic density and on the direct LSP detection 
rate, assuming the LSP to be higgsino--like. The relic density is in this
case often determined by co--annihilation processes, the rate of which
depends exponentially on the mass splittings between the higgsino--like
states. Yukawa loop corrections can therefore change the tree--level prediction
by a factor of $\sim 5$ in either direction. This re--introduces a state with
more than 99\% higgsino purity as a viable cold Dark Matter candidate, if
the corrections to the mass splittings are near the upper end of their allowed
range. If $\mu < 0$, the effect of loop corrections on the estimated LSP
counting rate is even more dramatic: The predicted rate might increase by
two orders of magnitude, but it might also be exactly zero (for spinless
nuclei), even if all sparticle masses are in or below the few hundred GeV
range. Clearly effects of this size have to be included in any quantitative
analysis of the properties of higgsino--like Dark Matter.

We conclude with some remarks regarding the viability of models with 
higgsino--like LSP. Within the framework of minimal supergravity models
\cite{43}, which assume universal scalar masses as well as unified gaugino
masses at the Grand Unification scale, a higgsino--like LSP is possible only
if $\tan^2 \beta \gg 1$, and if scalar soft breaking masses are significantly
larger than gaugino masses. Since gaugino masses in turn must be considerably
larger than $|\mu|$ for the LSP to be higgsino--like, naturalness arguments
favour a very light LSP in such a scenario. On the other hand, the
parameter space leading to a higgsino--like LSP opens up considerably if one
allows the sparticle spectrum at the GUT scale to be non--universal. In
particular, the predicted value of $|\mu|$ can be reduced either by giving
larger soft breaking masses to the Higgs bosons than to third generation
squarks, or by reducing the gluino mass compared to the masses of the
electroweak gauginos. We therefore conclude that a higgsino--like LSP with
mass slightly below $M_W$ can be a viable cold Dark Matter candidate, both
from the phenomenological and from the model building point of view.

\subsection*{Acknowledgements}
M.D. thanks the members of the Center for Theoretical Physics at Seoul
National University as well as the KEK theory group for their
hospitality during the course of this work. M. M. N. was supported in part by 
the Grant-in-aid for Scientific Research from the Ministry 
of Education, Science and Culture of Japan(07640428).

\section*{Appendix: Expressions for $C-$Functions}
\renewcommand{\theequation}{A.\arabic{equation}}
\setcounter{equation}{0}
In Sec.~2 we gave general expressions for the one--loop corrections from
Yukawa interactions to the $Z \lsp \lsp$ vertex, eqs.(12), as well as
for the $\phi \lsp \lsp$ and $A \lsp \lsp$ couplings, eqs.(16), in terms of
the $C-$functions defined in ref.\cite{14}. However, as already mentioned
in Sec.~2, the expressions for the $C-$functions contain apparent divergencies
both in the limit $s \rightarrow 4 \mchisq$ relevant for the calculation of
the LSP relic density, and in the limit $s \rightarrow 0$ relevant for the
calculation of the LSP--nucleon scattering cross section.\footnote{Recall
that eqs.(12) and (16) have been written in a convention where both momenta
$k_1$ and $k_2$ point towards the vertex. In case of LSP--nucleon scattering
the sign of one these momenta therefore has to be inverted.} In case of the
higher $C-$functions the necessary cancellations become too delicate for a
reliable numerical treatment even if ``double precision'' variables are used.
We have therefore re--evaluated the relevant Feynman parameter integrals in
these two kinematical limits, which allows us to express the $C-$functions
appearing in Sec.~2 as combinations of $B-$functions. These expressions are
collected in this Appendix.

In our notation the scalar three--point function $C_0$ is defined as
\beq \label{a1}
C_0(s,m^2,M_1,M_2,M_3) &= - \int_0^1 d y \int_0^y dx \left[
m^2 y^2 + s (x^2 - xy) + y \left(M_2^2 - M_3^2 - m^2 \right) 
\right. \nonumber \\ & \left. \hspace*{3cm} 
+ x (M_1^2 - M_2^2) + M_3^2 - i \epsilon \right]^{-1};
\eeq
this definition coincides with that used in Appendix C of ref.\cite{14}.
This gives:
\ben \label{a2} \beq
C_0(4m^2,m^2,M_1,M_2,M_3) &= \frac{1}{D} \left[ B_0(m^2,M_1,M_3) +
B_0(m^2,M_2,M_3) 
\right. \nonumber \\ & \left. \hspace*{.9cm} 
- 2 B_0(4 m^2,M_1,M_2) \right];
\label{a2a} \\
C_0(0,m^2,M_1,M_2,M_3) &= \frac {1} {M_1^2 - M_2^2} \left[ B_0(m^2,M_1,M_3)
- B_0(m_2,M_2,M_3) \right], \label{a2b}
\eeq \een
where
\be \label{a3}
D = 2 \left( m^2 + M_3^2 \right) - M_1^2 - M_2^2.
\ee
In the limit $M_1 \rightarrow M_2$, eq.(\ref{a2b}) reduces to:
\be \label{a4}
C_0(0,m^2,M,M,M_3) = - \frac {1} {2 m^2} \left[ \log \frac {M^2} {M_3^2}
+ \frac {M_3^2 + m^2 - M^2} {\sqrt{|\Delta|}} \cdot L \right],
\ee
where we have introduced
\ben \label{a5} \beq
\Delta &= 2 m^2 \left( M^2 + M_3^2 \right) - m^4 - \left(M^2 - M_3^2 \right)^2;
\label{a5a} \\
L &= \mbox{$ \left\{ \begin{array}{cc}
2 \arctan \frac {\sqrt{\Delta}} {M^2 + M_3^2 - m^2},
 \ \ \ \ \ & \Delta \geq 0\\
\log \frac {M^2 + M_3^2 - m^2 + \sqrt{-\Delta}} 
{M^2 + M_3^2 - m^2 - \sqrt{-\Delta}}, 
 \ \ \ \ \ & \Delta < 0
\end{array} \right. $}. \label{a5b}
\eeq \een

If eqs.(\ref{a2}) are used for $C_0$, the function $C_1^+$ defined in
ref.\cite{14} has apparent divergencies only at $s \rightarrow 4 m^2$
(we suppress the imaginary infinitesimal $-i \epsilon$ from now on):
\beq \label{a6}
C_1^+(4m^2,m^2,M_1,M_2,M_3) &= - \int_0^1 dy \int_0^y dx \frac
{y/2} {m^2 (y-2x)^2 + y \left(M_2^2 - M_3^2 - m^2 \right) + 
x \left(M_1^2 - M_2^2 \right) + M_3^2}
\nonumber \\
&= \frac {1}{2D} + \frac {1}{D^2} \left\{ M_3^2 \left[ B_0(m^2,M_1,M_3)
+ B_0(m^2,M_2,M_3) \right]
\right. \nonumber \\ & \left. \hspace*{2.3cm}
+ m^2 \left[ B_2(m^2,M_3,M_1) + B_2(m^2,M_3,M_2) \right] 
\right. \nonumber \\ & \left. \hspace*{2.3cm}
+ \frac {M_1^2 - M_2^2} {2} \left[ B_1(m^2,M_3,M_1) - B_1(m^2,M_3,M_2) \right]
\right. \nonumber \\ & \left. \hspace*{2.3cm}
- \left[ 2 \left( m^2 + M_3^2 \right) + M_2^2 - M_1^2 \right] B_0(4m^2,M_1,M_2)
\right.  \\ & \left. \hspace*{2.3cm}
+ 2 \left( M_2^2 - M_1^2 \right) B_1(4m^2,M_2,M_1) + 8 m^2 B_3(4m^2,M_1,M_2)
\right\}, \nonumber
\eeq
where $D$ has been defined in eq.(\ref{a3}). Here we have used the higher $B$
functions as defined in ref.\cite{44}; recall that our definition of $B_1$
differs by an overall sign from that of ref.\cite{14}.

Similarly, after application of eqs.(\ref{a2}), the function $C_1^-$ contains
apparent divergencies only at $s=0$:
\beq \label{a7}
C_1^-(0,m^2,M_1,M_2,M_3) &= - \int_0^1 dy \int_0^y dx \frac {x-y/2}
{m^2 y^2 + y \left( M_2^2 -M_3^2 - m^2 \right) + x \left( M_1^2 - M_3^2 
\right) + M_3^2}
\nonumber \\
&= \frac {1}{2 \left( M_1^2 - M_2^2 \right)} \left\{ \left( 1 + 2 \frac
{M_2^2 - M_3^2} {M_1^2 - M_2^2} \right) \left[ B_1(m^2,M_3,M_2)
- B_1(m^2,M_3,M_1) \right]
\right. \nonumber \\ & \left. \hspace*{2.9cm}
- \frac {2m^2} {M_1^2 - M_2^2} \left[ B_3(m^2,M_3,M_2) - B_3(m^2,M_3,M_1)
\right] \right. \\ & \left. \hspace*{2.9cm}
+ \frac {2M_3^2} {M_1^2 - M_2^2} \left[ B_0(m^2,M_2,M_3) - B_0(m^2,M_1,M_3)
\right] - 1 \right\}. \nonumber
\eeq
Note that by construction \cite{14}, $C_1^- \rightarrow 0$ as $M_1 \rightarrow
M_2$.

The functions $C_2^+$ and $C_2^-$ only appear in the $Z$ vertex, eqs.(12).
Moreover, the coefficient in front of $C_2^+$ vanishes for $s \rightarrow
4 \mchisq$, while the coefficient in front of $C_2^-$ vanishes for $s 
\rightarrow 0$. We therefore only need to consider $C_2^+$ in the limit
$s \rightarrow 0$:
\beq \label{a8}
C_2^+(0,m^2,M_1,M_2,M_3) &= - \int_0^1 dy \int_0^y dx \frac {y^2/4}
{m^2 y^2 + y \left( M_2^2 -M_3^2 - m^2 \right) + x \left( M_1^2 - M_3^2 
\right) + M_3^2}
\nonumber \\
&= \frac {1} {4 \left( M_1^2 - M_2^2 \right)} \left[ B_2(m^2,M_3,M_2) -
B_2(m^2,M_3,M_1) \right].
\eeq
In the limit $M_1 \rightarrow M_2$, this reduces to
\be \label{a9}
C_2^+(0,m^2,M,M,M_3) = \frac{1}{4} \left[ B_1'(m^2,M_3,M) + B_0'(m^2,M_3,M)
+ C_0(0,m^2,M,M,M_3) \right],
\ee
where $C_0(0,m^2,M,M,M_3)$ is given in eq.(\ref{a4}), and $B_0'$ and $B_1'$
are the derivatives of $B_0$ and $B_1$ with respect to their first 
argument.\footnote{Of course, eq.(\ref{a9}) can also be written as derivative
of $B_2$ with respect to its {\em last} argument, since all two-- and
three--point functions only depend on the squares of the masses appearing as
arguments. However, conventionally one only uses derivatives with respect to
the first argument.} Similarly, we need $C_2^-$ only in the limit $s
\rightarrow 4 m^2$:
\beq \label{a10}
C_2^-(4m^2,m^2,M_1,M_2,M_3) &= - \int_0^1 dy \int_0^y dx \frac {(x-y/2)^2}
{m^2 (y-2x)^2 + y \left(M_2^2 - M_3^2 - m^2 \right) + 
x \left(M_1^2 - M_2^2 \right) + M_3^2}
\nonumber \\
&= \frac{1}{4D} \left[ B_2(m^2,M_3,M_1) + B_2(m^2,M_3,M_2)
\right. \nonumber \\ & \left. \hspace*{1.1cm} 
+ 8 B_3(4m^2,M_1,M_2) - 2 B_0 (4 m^2, M_1, M_2) \right],
\eeq
where $D$ is again given by eq.(\ref{a3}).

Finally, we note that the divergent function $C_2^0$ that appears in eqs.(12)
can be computed from the general expression given in eq.(C4) of 
ref.\cite{14}, using the results for $C_0, \ C_1^+$ and $C_1^-$ collected in
this Appendix.

\newpage
\section*{Figure Captions}

\vspace{5mm}
\noindent
{\bf Fig.~2:} The chargino--LSP mass difference (solid), the 
axial--vector $Z \lsp \lsp$ coupling (long dashed), and the $h^0 \lsp \lsp$
coupling (short dashed) as a function of the soft breaking $A$
parameter, including one--loop corrections involving Yukawa couplings.
``Low'' and ``high'' leading--order estimates for the 
branching ratio for inclusive radiative $b$ decays are also shown; 
for our assumption of
exactly universal weak--scale soft breaking squark masses, this gives the
strongest constraints on $A$ in the region of interest, as discussed in the
text. All quantities have been re--scaled, as indicated.

\vspace{5mm}
\noindent
{\bf Fig.~3:} The chargino--LSP mass difference (a), the LSP
relic density \Ochi\ (b), and the expected LSP detection rate in a
${}^{76}$Ge detector (c), as a function of the gaugino fraction,
defined as the sum of the squares of the gaugino components of the LSP
eigenvector. These results are for a fixed LSP mass, so that both
$M_2$ and $\mu$ vary along the text.  Further, $|A|$ has been
decreased from $2.7 m_{\tilde q}$ to $2.5 m_{\tilde q}$ as $M_2$ was
increased from about 150 GeV to 1 TeV.

\vspace{5mm}
\noindent
{\bf Fig.~4:} As in Fig.~3, but for positive sign of $\mu$; also, $|A|$ has
been kept fixed in this figure, and \mchi\ has been reduced by 2 GeV, in
order to be closer to the region of parameter space where the prediction for
\Ochi\ is maximized.

\vspace{5mm}
\noindent
{\bf Fig.~5:} In the region of the $(\mu<0, M_2)$ half--plane corresponding to a
higgsino--like LSP, we show contours of constant \Ochi\ (dashed) and 
contours of constant LSP detection rate(CR) in a ${}^{76}$Ge detector, 
in units of events/(Kg $\cdot$ day) (dotted and
dot--dashed); note that the values on the latter contours are different for
Figs. (a) (no Yukawa loop corrections), (b) ($A<0$) and (c) ($A>0$). 
The region to the right of the solid line is excluded by our
interpretation of the LEP chargino search limit, as discussed in the text.

\end{document}